\documentclass[prl,reprint,noeprint,superscriptaddress,footinbib,aps]{revtex4-2}
\usepackage[nolist]{acronym}
\usepackage{siunitx}
\usepackage{physics}
\usepackage{miller}
\usepackage[colorlinks=true,linkcolor=blue,anchorcolor=blue,citecolor=blue,filecolor=black,menucolor=blue,runcolor=blue,urlcolor=blue]{hyperref}
\usepackage{graphicx}

\DeclareSIUnit{\PU}{PU}
\DeclareSIUnit{\PUs}{PU}
\DeclareSIUnit\angstrom{\protect \text {Å}}
\AtBeginDocument{\RenewCommandCopy\qty\SI}

\newcommand{\gnrs}{$\left(3,2,8\right)$-chGNRs}

\newcommand{\gdau}{$\textrm{GdAu}_2$}
\newcommand{\au}{Au\hkl(111)}
\newcommand{\beff}{$B_\mathrm{eff}$}
\newcommand{\Ueff}{$U_\mathrm{eff}$}
\newcommand{\Bc}{$B_\mathrm{eff}^\mathrm{c}$}
\newcommand{\muL}{$\Delta\mu_\mathrm{L}$}
\newcommand{\muc}{$\Delta\mu_\mathrm{L}^\mathrm{c}$}

\newcommand{\cru}{\ensuremath{\hat{c}_{\mathrm{R}\uparrow}^\dagger}}

\newcommand{\crd}{\ensuremath{\hat{c}_{\mathrm{R}\downarrow}^\dagger}}
\newcommand{\vac}{\ensuremath{\ket{0}}}

\newcommand*{\subfigref}[2][]{%
	\hyperref[{#2}]{%
		Figure~\ref*{#2}%
		\ifx\\#1\\%
		\else
		#1%
		\fi
	}%
}

\begin{document}

\title{Spin and Charge Control of Topological End States in Chiral Graphene Nanoribbons on a 2D Ferromagnet}

\author{L. Edens}
\author{F. Romero Lara}
\author{T. Sai}
\author{K. Biswas}
\affiliation{CIC nanoGUNE BRTA, 20018 San Sebastián, Spain}

\author{M. Vilas-Varela}
\affiliation{Centro Singular de Investigaci\'on en Qu\'imica Biol\'oxica e Materiais Moleculares (CiQUS) and Departamento de Qu\'imica Org\'anica, Universidade de Santiago de Compostela, 15782 Santiago de Compostela, Spain}

\author{F. Schulz}
\affiliation{CIC nanoGUNE BRTA, 20018 San Sebastián, Spain}

\author{D. Peña}
\affiliation{Centro Singular de Investigaci\'on en Qu\'imica Biol\'oxica e Materiais Moleculares (CiQUS) and Departamento de Qu\'imica Org\'anica, Universidade de Santiago de Compostela, 15782 Santiago de Compostela, Spain}
\affiliation{Oportunius, Galician Innovation Agency (GAIN)
15702 Santiago de Compostela, Spain}

\author{J. I. Pascual}
\affiliation{CIC nanoGUNE BRTA, 20018 San Sebastián, Spain}
\affiliation{Ikerbasque, Basque Foundation for Science, 48013 Bilbao, Spain}

\date{\today}

\begin{abstract}
Tailor-made graphene nanostructures can exhibit symmetry-protected topological boundary states that host localized spin-$1/2$ moments. However, one frequently observes charge transfer on coinage metal substrates, which results in spinless closed-shell configurations. Using low temperature scanning tunneling spectroscopy, we demonstrate here that pristine topologically nontrivial chiral graphene nanoribbons synthesized directly on the ferromagnet \gdau{} can either maintain a charge-neutral diradical singlet or triplet configuration, or exist in a singly anionic doublet state. As an underlying mechanism, we identify a moir\'{e}-modulated work function and exchange field, as corroborated by Kelvin-probe force microscopy and spin-flip spectroscopy. The joint electrostatic and magnetic interactions allow reversibly switching between the three spin multiplicities by atomic manipulation. We introduce an effective Hubbard dimer model that unifies the effects of local electrostatic gating, electron-electron-correlation, hybridization and exchange field to outline the phase diagram of accessible spin states. Our results establish a platform for the local control of $\pi$-radicals adsorbed on metallic substrates.
\end{abstract}
\maketitle
\begin{acronym}
    \acro{chGNR}{chiral graphene nanoribbon}
    \acro{AFM}{atomic force microscope}
    \acro{STM}{scanning tunneling microscope}
    \acro{MFH}{mean-field Hubbard model}
    \acro{SPTES}{symmetry-protected topological end state}
    \acroplural{SPTES}[SPTES]{symmetry-protected topological end states}
    
    \acro{HOMO}{highest occupied molecular orbital}
    \acro{SOMO}{singly occupied molecular orbital}
    \acro{SUMO}{singly unoccupied molecular orbital}
    \acro{LDOS}{local density of states}
    \acro{PU}[PU]{precursor unit}

    \acro{OSS}{on-surface synthesis}    \acro{KPFM}{Kelvin probe force microscopy}
    \acro{LCPD}{local contact potential difference}
\end{acronym}

\section{Introduction}
Despite its many remarkable properties, graphene is intrinsically diamagnetic in bulk form. Yet, in finite geometries, local magnetic moments can form due to the alignment of unpaired $p_z$ electrons along boundaries \cite{DeCar22}. In particular, such nanographenes can form high-spin configurations through mutual exchange interaction between localized zero-energy states \cite{Li2020a,HieOnS21,VilOnS23,VegOns25}. Graphene nanostructures in nearly arbitrary shapes can be synthesized from from specifically designed precursors via \ac{OSS} on noble metals under ultra-high vacuum conditions \cite{CaiAto10,ClaCon19}.

\begin{figure*}
	\includegraphics[scale=1]{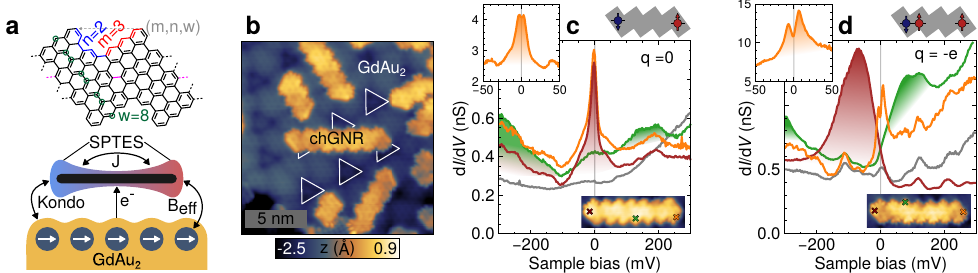}
	\caption{Probing \gnrs{} on \gdau{}. \textbf{(a)} The chemical structure, and a scheme of the competing interactions between topological end states and the ferromagnetic surface. \textbf{(b)} \acs*{STM} scan of \acsp*{chGNR} (\SI{0.4}{\V}, \SI{1}{\nA}, \qtyproduct{15x15}{\nm}). \textbf{(c, d)} Differential conductance spectra of two \SI{7}{\PU}-\acsp*{chGNR} on different adsorption sites. Insets shows tip positions where spectra were recorded. Outset shows high resolution spectra of the split Kondo feature in a narrow energy range. The former is found in a fully neutral charge state, with two screened radical moments. The latter exists in a singly anionic doublet configuration. Spectra without Kondo resonances have been scaled and offset for better comparison.}
	\label{fig-sts}
\end{figure*}

The open-shell characteristic in nanographene can be understood as a consequence of sublattice imbalance \cite{FerMag07}, topological frustration \cite{WanTop09}, or bulk-boundary correspondence \cite{CaoTop17}. The latter is at play in \acp{chGNR}, whose translation vector $(m,n)$ lies between the primitive armchair and zigzag directions as drawn in \subfigref[a]{fig-sts} \cite{NakEdg96a,YazThe11,DeSub16,MerUnr18}. Due to their finite width $w$, the bands formed along their chiral edges overlap and introduce a spectral gap which can be engineered to be nontrivial by appropriate choice of $(m,n,w)$ \cite{Li2021}. In these special cases, two zero-energy modes confined to the two termini emerge in finite-length \acp{chGNR}. Owing to their single occupancy at half filling, the resulting radical system can be considered as an interacting pair of electron spins. Given the intrinsically balanced sublattices, Lieb's theorem predicts these \acp{SPTES} to align antiferromagnetically to form an overall $S=0$ singlet state \cite{Li2021}.

The localized magnetic character of \acp{SPTES} in \acp{chGNR} has so far eluded  experimental observation due to work function mismatches between graphene and common catalyst substrates \cite{WanAza22}. On the electronegative surface \au{}, \acp{chGNR} undergo significant cationic charge transfer \cite{Li2021}, while more electropositive supports like Ag\hkl(111), Ag\hkl(100) and MgO/Ag\hkl(100) give highly anionic states \cite{CorBan21,DomEng24}. In these cases, both \acp{SPTES} are fully (de)populated, leading to a nonmagnetic closed-shell state.  We demonstrate here that the catalytically active intermetallic \gdau{} \cite{CorAu110,QueTwo20,QueSur20,BreDet23} supports perfect charge neutrality in \gnrs{} due to an intermediate work function, allowing us to sense the spin localized to each terminus.

\gdau{} is also known to display 2D in-plane ferromagnetic order at low temperatures \cite{OrmHig16,BazAto19,BreDet23}. The interaction between polyradical boundary states in nanographene and a ferromagnetic substrate however remains largely unexplored. In this work, we spectroscopically probe the significant exchange fields \cite{PasKon04,HauEle08,GaaUni11,MisUnd12,NisExc15,YanCon15,YanTun19,BasSen25} imposed by the substrate onto $\pi$-radicals and study its effect on the molecular spin multiplicity of \acp{chGNR}. Spin-flip spectroscopy and many-body modeling confirm magnetic coupling to the \gdau{} can dominate over the intramolecular exchange between \acp{SPTES}. As a result, a well-separated high-spin configuration is demonstrated to form in directly adsorbed \acp{chGNR} that would otherwise lack a defined magnetic state.

The unique moir\'{e} superstructure stemming from the angled alignment between the \gdau{} overlayer and \au{} opens further possibilities of controlling spin: the exchange field depends strongly on the adsorption position, which enables us to tune or even nullify it entirely by reversible lateral manipulation of individual \acp{chGNR}. Additionally, we encounter configurations where one electron charge is transferred to a single \ac{SPTES}, thereby allowing us to also perform local electrostatic gating by controlled reorientation. Even with one of the \acp{SPTES} radicals paired, and thereby in the absence of intramolecular exchange, the remaining doublet still remains well defined by the magnetic coupling to the substrate alone.

\begin{figure*} [t!]
	\includegraphics[scale=1]{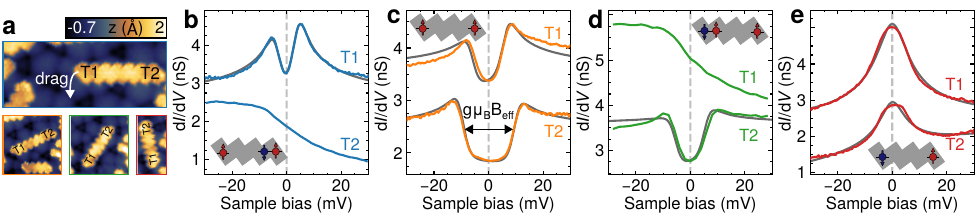}
	\caption{Controlling the spin multiplicity of a \SI{7}{\PU} \ac{chGNR} on \gdau{}. By lateral manipulation of the same molecule into four different positions, we observe the excess charge leaving terminus T2 \textbf{(b)} and re-entering T1 \textbf{(d)}. The intermediate neutral case is realized in \textbf{(c)} with large a large exchange field. In the last position \textbf{(e)}, \beff{} is smaller than the Kondo coupling on either termini and the zero-bias resonance is restored. Fits of the inelastic tunneling  theory \textbf{(b-d)} and Frota lineshapes \textbf{(e)} are shown in gray. Traces have been offset and scaled for clarity.}
	\label{fig-switch}
\end{figure*}

\section{Results and Discussion}

Following \ac{OSS} on \gdau{}, \acfp{chGNR} naturally align with respect to the underlying moir\'{e}-induced superstructure, as apparent in the \ac{STM} image in \subfigref[b]{fig-sts}. Bond-resolved \ac{AFM} imaging confirms their defect-free chemical structure \footnote{Supporting information containing references \cite{SanHub21,CheChe76,JafInt08}}. As seen in \subfigref[c]{fig-sts} for a 7 \acp{PU} long \acp{chGNR}, differential conductance tunneling spectra recorded on the termini reveal two sharp zero-bias features, whereas the \ac{LDOS} on the chiral edge only shows the onsets of characteristic edge bands. We attribute the former features to Kondo resonances arising from separate screening of each $\pi$-radical by conduction electrons. Their emergence indicates a charge-neutral, half-filled \ac{chGNR}, as corroborated by the symmetric position of the Fermi level within the large gap.

High-resolution spectra of the zero-bias feature often reveal a small splitting thereof, which suggests a magnetic coupling; its origin will be addressed below. The spectra in \subfigref[c]{fig-sts} also display two lower peaks straddling the Kondo resonance near-symmetrically at \SI{\pm40}{\mV}, which we assign to the singly (un)occupied \acp{SPTES} zero modes split off by intramolecular Coulomb repulsion \footnotemark[1]. The presence of such a correlation gap is yet another indication of charge neutrality and the stabilization an open-shell configuration. 

We observe a second class of otherwise identical \acp{chGNR} in other adsorption geometries. For instance, in \subfigref[d]{fig-sts}, a (split) Kondo feature is encountered only at one of the termini, whereas the other displays a much broader ($>\!\!\!\SI{100}{\mV}$) resonance well below the Fermi level. We attribute this second feature to a doubly populated \ac{SPTES}, which implies a singly anionic net charge ($q=-e$). In a single-particle picture, the surplus electron here resides within the left terminus and pairs with its $\pi$-radical, forming the \ac{HOMO}. Negative charge accumulation is consistent with the Fermi level now residing closer towards the conduction band, which remains localized along the chiral edge. As confirmed by the \ac{MFH} calculations, this species forms a $S=1/2$ ground state, which we refer to as the charged doublet \footnotemark[1]. 

Further insight into the origin of two coexisting charge states is gained by repeated manipulation of a single \ac{chGNR} into several adsorption positions with the \ac{STM} tip. For instance, the naturally adsorbed 7 \ac{PU} \ac{chGNR} presented in \subfigref[a,b]{fig-switch} shows a doubly occupied \ac{SPTES} on terminus T2, and a split Kondo resonance on the other. When rotated slightly, the molecule becomes charge-neutral, with split Kondo resonances on both termini. After repositioning by a few nanometers, an excess electron is gained by T1, now displaying a \ac{HOMO}, whereas the local moment of T2 remains. In the final position (\subfigref[e]{fig-switch}), a pair of zero-bias peaks with no resolvable splitting is restored, again indicating perfect half-filling.

The reversible orientation-dependent switching between the anionic and neutral charge states suggests a varying degree of electrostatic interaction with \gdau{}. To explore its local character, we map the \ac{LCPD} across the moir\'{e} superstructure using \ac{KPFM}. As seen in \subfigref[a]{fig-kpfm}, the triangular moir\'{e} minima possess a decreased \ac{LCPD} over the rest of the unit cell by approximately \SI{80}{\milli\eV}, indicating a locally smaller substrate work function, and hence an electron-donating character \cite{QueTwo20}. In lateral manipulation experiments, these features therefore act as a positive local gating potential \cite{MarGat15}: when a \ac{chGNR} terminus lies nearby, a single electron is transferred, giving double occupancy of one \acp{SPTES}.

\ac{KPFM} has also been applied to map charge distribution down to the intramolecular level \cite{MohIma12,SchCon14,AlbPro15,JelHig17}. Here, we sense the spatial extent of the excess electron residing in terminus T1 of the charged doublet \ac{chGNR} shown in \subfigref[b]{fig-kpfm}. The \ac{KPFM} image (\subfigref[c]{fig-kpfm}) reveals two key features. For one, the pillow effect the molecule has on \gdau{} results in a decreased work function ($\approx\SI{-100}{\meV}$) in a wide perimeter around its boundary \cite{CriCha02}. Second, we note a sharp increase in \ac{LCPD} in the upper section of T1, which we assign to the expected negative charge density. Terminus T2, with a singly occupied \ac{SPTES} \footnotemark[1], exhibits no such feature. Remarkably, the electrostatic potential emanating from T1 is large enough to negate both the pillow effect, and the smaller work function of the moir\'{e} minimum underneath. 

As is evident in \autoref{fig-switch}, the magnitude of the inelastic gap making up the split Kondo feature also varies with adsorption geometry. We attribute this effect to a locally varying exchange field exerted onto \acp{chGNR} by \gdau{}, a known 2D in-plane ferromagnet at low temperatures \cite{OrmHig16,BazAto19,BreDet23}. Due to their proximity, the radicals thereby experience an effective magnetic field, which can be understood as the consequence of spin-dependent Kondo hybridization proportional to the substrate electrons' density and spin polarization \cite{PasKon04,HeeKon06,ChoKon16,BasSen25}. From fits of a third-order inelastic tunneling model \cite{TerSpi15}, assuming a free electron $g$-factor, we obtain \beff{} to be as large as \SI{84}{\tesla} on T2 in \subfigref[c]{fig-switch}.

The exchange field often differs significantly in magnitude even between termini, hinting at a magnetic substrate coupling modulated on a similar scale to the work function. Presumably, $\pi$-radicals interact with several $\textrm{Gd}^{3+}$ $4f$ moments via an indirect Au $5\mathrm{d}$-mediated mechanism, which has already been suggested as the coupling mechanism responsible for the in-plane magnetic order in \gdau{} \cite{OrmHig16}. The resulting conduction electron polarization was shown to exhibit a texture congruent to the moir\'{e} superstructure, consistent with our observations \cite{BazAto19,BreDet23}. Additionally, since the Gd sublattice is spaced relatively far apart at \SI{5.5}{\angstrom} \cite{CorAu110}, we expect stronger exchange in lateral placements where the number of sites covered by each \ac{chGNR} is maximal.

\begin{figure*}
	\includegraphics[scale=1]{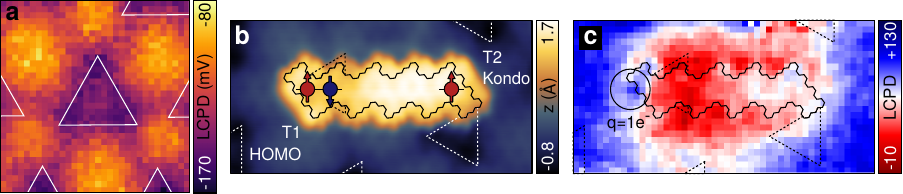}
	\caption{Visualizing the localized charging effect.  \textbf{(a)} \acs*{KPFM} of bare \gdau{} reveals a decreased local work function in the triangular moir\'{e} minima by tens of \unit{\meV} (junction resistance \SI{140}{\mega\ohm}, oscillation amplitude \SI{50}{\pm}, \qtyproduct{5x5}{\nm}). \textbf{(b)} The shown \acs*{chGNR} exhibits a doubly occupied orbital on terminus T1, laying within a moir\'{e} minimum (\SI{0.5}{\V}, \SI{1}{\nA}, \qtyproduct{9x4.5}{\nm}). \textbf{(c)} The left \acs*{SPTES} shows a localized increase in \acs*{LCPD}, consistent with a single excess electron residing in T1 (\SI{865}{\Mohm}, \SI{50}{\pm}, \qtyproduct{9x4.5}{\nm}).}
	\label{fig-kpfm}
\end{figure*}

In the absence of \beff{}, intramolecular Coulomb repulsion promotes antiferromagnetic alignment between $\pi$-radicals in charge-neutral \acp{chGNR}, especially for short lengths where \acp{SPTES} overlap and hybridize. When interfaced with \gdau{}, each spin residing at each terminus is additionally exchange-coupled to to the magnetic surface. We parametrize these competing interactions as $J>0$, giving antiferromagnetic alignment, and \beff{}, here assumed identical for both termini. If \beff{} is smaller than $J$, the Kondo-screened singlet state persists and no substrate-induced inelastic conductance gap is resolved, as in \subfigref[e]{fig-switch}.
If, however, \beff{} exceeds $J$, a triplet state is favored over the singlet state, in which both radicals align with the substrate magnetization and spin-flip excitations emerge. 

Due to the extended nature of nanoribbons, intramolecular exchange is only a secondary magnetic coupling in our experiment: in molecules with $L\geq\SI{5}{\PU}$, a zero-bias resonance is consistently restored by proper reorientation on the substrate. $J/k_\mathrm{B}$ therefore must be smaller than the Kondo temperature $T_\mathrm{K}$ which can be calculated \cite{TurDem24} to amount to \SI{23}{\K} on T2 in \subfigref[d]{fig-switch}. Under the influence of Kondo scattering, the singlet ground state is therefore not well defined for long \acp{chGNR}. The triplet state, on the other hand, is stable against these fluctuations if $\abs{g\mu_\mathrm{B}B_\mathrm{eff}}>k_\mathrm{B}T_\mathrm{K}, J$ \cite{SpiExp15}. In this way, the built-in exchange field protects $\pi$-radical moments from scattering, as evidenced by an inelastic gap,  despite being in contact with a fully metallic surface \cite{LotBis12}.

The crossover between charge-neutral singlet and triplet, as well as that towards the charged doublet state is governed by several effects: $\pi$-delocalization and electron-electron repulsion intrinsic to the molecule, as well as the local gating and exchange field imposed by \gdau{}.
To capture their intricate interplay in stabilizing each spin state, we map \acp{chGNR} onto a modified Hubbard dimer model, in which each site represents one \ac{SPTES} distributed across several graphene lattice sites as illustrated in \subfigref[a]{fig-dimer} \cite{OrtEle18a,DeCar22}. As becomes apparent from a \ac{MFH} description of the full $\pi$-lattice, our dimer approximation is valid in the limit of small to intermediate on-carbon-site repulsion $U_\mathrm{H}<\SI{3}{\eV}$ and charge states between $-2$ and $2$ electrons. Within this range, the spin-polarized active space is restricted to the \acp{SPTES}, which remain energetically well isolated from the quantized edge-localized bands \footnotemark[1]. 
We formulate our description in the grand canonical ensemble, allowing independent gating of each site via site-specific chemical potentials, and separate magnetic fields. The resulting Hamiltonian, expressed in the second quantization formalism, reads:
\begin{equation}
	\begin{gathered}
		\hat{\mathcal{H}}=-t\sum_{i\neq j}\sum_\sigma \hat{c}_{i\sigma}^{\dagger}\hat{c}_{j\sigma}+\sum_i U_i \hat{n}_{i\uparrow}\hat{n}_{i\downarrow}\\
		-\sum_i\sum_\sigma \mu_i \hat{n}_{i\sigma}+g\mu_\mathrm{B}\sum_i B_i \left(\hat{n}_{i\uparrow}-\hat{n}_{i\downarrow}\right).
	\end{gathered}
    \label{eq-hubdim}
\end{equation}
The first row corresponds to the conventional Hubbard dimer, with $i,j$ being the site indices L and R, a hopping amplitude $t$ representing length-dependent hybridization between \acp{SPTES}, and an effective Coulomb repulsion $U_i$ among end-localized $\pi$-electrons of spin $\sigma$. The chemical potentials $\mu_i$ and the Zeeman effect enter as additional diagonal elements, for which we set $g=-2$. Here, we assume identical repulsion $U_i=U_\textrm{eff}$ and neutrality of site $\mathrm{R}$, such that $\mu_\mathrm{R}=U_\textrm{eff}/2$ and impose a symmetric exchange field $B_i=B_\textrm{eff}$. Lastly, we introduce $\Delta\mu_\mathrm{L}=\mu_\mathrm{L}-U_\textrm{eff}/2$ as the deviation of the left site from half filling, accounting in this way for the work function variations found on \gdau{}. The general, fully asymmetric case is discussed in \footnotemark[1].

\begin{figure*}
	\includegraphics[scale=1]{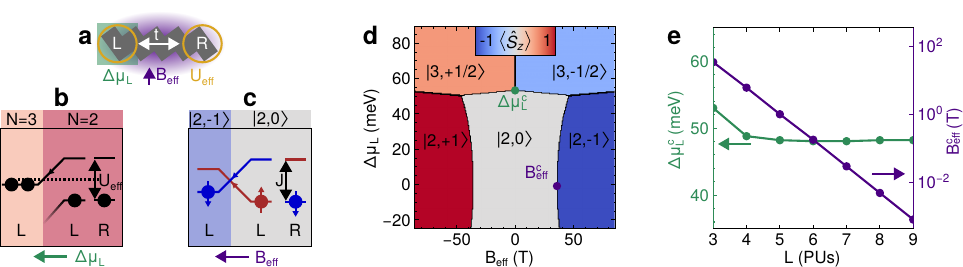}
	\caption{Predicting the \acs*{chGNR} spin multiplicity from the Hubbard dimer model. \textbf{(a)} Sites L and R, representing \acp{SPTES} under an exchange field \beff{}, experience Coulomb repulsions \Ueff{} and are coupled by a hopping amplitude $t$, both obtained from tight-binding theory of \gnrs{}. Each site is additionally gated by $\mu_\mathrm{L}$ and $\mu_\mathrm{R}$, respectively. The many-body eigenstates $\ket{N,m_s}$ evolve with the local potential $\Delta\mu_\mathrm{L}=\mu_\mathrm{L}-U_\textrm{eff}$ and \beff{}. \textbf{(b)}  Ground-state transition from two-particle states towards the charged doublet $\ket{3,\pm1/2}$ with \muL{}. \textbf{(c)} Spin level scheme of the singlet-triplet transition at a critical value of \beff{}. Only one component of $\ket{2,0}$ is shown. 
    \textbf{(d)} The \ac{chGNR} phase diagram at $\mu_\mathrm{R}=U_\textrm{eff}/2$, here for $L=3$, exhibits two critical points in \muL{} and $B_\textrm{eff}$. Color denotes the total spin projection $\langle S_z \rangle$ in units of $\hbar$. \textbf{(e)} \muc{} and \Bc{} as a function of length. The intramolecular exchange $J=\abs{g\mu_\mathrm{B}B_\textrm{eff}^\mathrm{c}}\approx 4t^2/U_\textrm{eff}$ falls rapidly, while the critical gating potential approaches $U_\textrm{eff}/2$.}
	\label{fig-dimer}
\end{figure*}

Exact diagonalization yields solutions we label as $\ket{N,m_s}$, where $N=2, 3$ is the \ac{SPTES} occupancy found in our experiment and $m_s$ the total spin projection. These are, in general, entangled combinations of single-particle states such as $\cru{}\crd{}\vac{}$ sketched in the insets of the experimental figures above. Depending on \muL{}, $B_\textrm{eff}$ and $t$, five many-body solutions can constitute the ground state. At half filling ($N=2$), the system may reside either in a neutral singlet state $\ket{2,0}$  or in one of the triplet states  $\ket{2,\pm 1}$. 
For sufficiently long \acp{chGNR}, such that $U_\textrm{eff}\gg t$, the singlet approximately consists in a superposition of two singly occupied \acp{SPTES} as  -- an open-shell configuration. The triplet states $\ket{2,\pm 1}$ represent ferromagnetically aligned \acp{SPTES}, and become non-degenerate under the exchange field.
For $N=3$, we identify the charged doublet $\ket{3,\pm 1/2}$, also split by $B_\textrm{eff}$. Its energetic separation from the two-electron states is primarily given by the energy gain associated to site L via the local gating potential.

To assess the magnetic ground state of the \acp{chGNR} in our experiment using the Hubbard dimer, we first calculate effective parameters from the tight-binding description of the full $\pi$-lattice over a range of lengths $L$. The dimer hopping amplitude is extracted from the small hybridization gap between \acp{SPTES}, which amounts to $2t$, and decays exponentially with their separation. The nearly length-independent effective repulsion \Ueff{} is equal to the Coulomb integral over each \ac{SPTES} scaled by  $U_\mathrm{H}=\SI{3}{\eV}$, and bears only a small correction due to orbital overlap \footnotemark[1]. The number of \acp{PU} thereby directly controls the ratio $U_\textrm{eff}/t$, key measure of electron localization. 

The ground-state phase diagram spanned by the parameters \muL{} and \beff{} is best visualized at short lengths, where the singlet $\ket{2,0}$ is most stable; in \subfigref[d]{fig-dimer}, we set $L=\SI{3}{\PU}$. Along the vertical axis, we identify the critical point \muc{}, which marks the boundary between neutral singlet and charged doublet ground states. Its origin is most apparent by considering the occupation energy levels as depicted in \subfigref[b]{fig-dimer}: the required double population of site L becomes favorable only for $\Delta \mu_\mathrm{L}>U_\mathrm{eff}/2$. If \muL{} is large, these solutions simply correspond to a doubly occupied site L and a half-filled site R, i.e. a configuration with one unpaired electron on one terminus. 

Similarly, \Bc{} marks the transition between singlet and triplet states. As exemplified by the spin level scheme in \subfigref[c]{fig-dimer} drawn for a single-particle state, spin down (up) levels in the composition of $\ket{2,0}$ are shifted down (up) in energy by $B_\mathrm{eff}>0$, resulting in the ferromagnetically aligned $\ket{2,-1}$. Due to hybridization, the magnitudes of \Bc{} and \muc{} generally depend on $U_\textrm{eff}/t$, and thereby on $L$. For instance, increasing the length from \qtyrange{3}{9}{\PU}, the critical field \Bc{} decreases from \SI{34}{\tesla} to below \SI{1}{\milli\tesla}, as displayed in \subfigref[e]{fig-dimer}. If \muL{} is negligible, the description simplifies to a Heisenberg-coupled spin dimer in the limit of large $L$, with a singlet-triplet gap amounting to $J=\abs{g\mu_\mathrm{B}B_\textrm{eff}^\mathrm{c}}=4t^2/U_\textrm{eff}$ \cite{OrtEle18a,JacThe22,DeCar22}.

Based on the small magnetic coupling between termini, we hypothesize that the neutral \SI{7}{\PU}-\ac{chGNR} in \subfigref[c]{fig-switch} resides in a stable triplet ground state, as the exchange field extracted from the inelastic gap exceeds both \Bc{} and the Kondo coupling energy by far. The observed spin-flip excitations are therefore from the triplet to the singlet configuration. Due to the localized nature of the \ac{SPTES}, this change in equilibrium magnetic configuration does not significantly modify the remaining molecular orbitals \footnotemark[1].

Conversely, as is apparent in \subfigref[e]{fig-dimer}, the critical gating potential $\Delta\mu_\mathrm{L}^{c}$ is only weakly length-dependent and tends to a constant value of $U_\textrm{eff}/2=\SI{48}{\milli\eV}$ for $L>\SI{5}{\PU}$. In this limit, the hybridization $t$ becomes insignificant, and \muL{} must only overcome Coulomb repulsion to doubly occupy the left site. The decreased work function detected within the moir\'{e} minima of \gdau{} is therefore sufficient in magnitude to induce the charged doublet state for all nanoribbon lengths realized here.

From the Kondo temperature determined above, we expect a singlet ground state stabilized purely by intramolecular exchange as opposed to \beff{}, namely $J>k_\mathrm{B}T_\mathrm{K}$, only for $L\leq\SI{3}{\PU}$. However, in this limit, \ac{SPTES} overlap grows excessive, and the $\pi$-radical state is compromised \cite{Li2021}. Split Kondo resonances are therefore most likely not observable in \acp{chGNR} on common non-magnetic substrates, where metallic screening reduces electron-electron correlations \cite{WanGia16a}.

\section{Conclusions}

\acresetall
In summary, we have investigated the charge and magnetic ground state of $(3,2,8)$ \acp{chGNR} synthesized on \gdau{} via \acl{OSS}. These exhibit a topologically non-trivial band structure, resulting in two \acp{SPTES} localized at their termini, which constitute a pair of antiferromagnetically exchange-coupled zero modes. Unlike conventional metallic substrates, \gdau{} maintains the \acp{chGNR} in a charge-neutral state, evidenced by the single occupancy of both end states, which we detect as two localized Kondo features in tunneling spectra. In some cases, we also encounter a negatively charged state, with a single excess electron hosted by one \ac{SPTES}, locally quenching spin. Uniquely, the in-plane magnetization of the rare-earth alloy induces an additional exchange field \beff{} that splits the Kondo resonances, providing a method to probe the local magnetic environment of the molecules. Interestingly, both the charge state and \beff{} can vary depending on the adsorption site, as confirmed by reversible molecular manipulation. The former is ascribed to local work function variations on the order of $\approx\SI{80}{\meV}$; similarly, the latter exhibits oscillations of up to several tens of Teslas. We attribute both effects to the moir\'{e}superstructure formed by the topmost atomic layer of \gdau{}. 

Using an asymmetric Hubbard dimer model, we show that the substrate-induced magnetic coupling can overcome the intrinsic exchange between end states, promoting a transition to a triplet ground state, where both spins at the termini are forced to lie parallel. The same model also shows the observed singly anionic state in terms of local electrostatic gating to be compatible with the work function modulation found in our experiment. The sensitivity of nanographene zero modes to \unit{meV}-range electrostatic backgrounds suggests that other rare-earth surface alloys with slightly different work functions could be used to selectively stabilize alternative charge states, including $+1$ and $-2$, which may exhibit novel magnetic states. 
Overall, our findings establish \gdau{} as a powerful platform for preserving and controlling the spin states of open-shell graphene nanostructures.
 
\section{Experimental section}
A pristine Au\hkl(111) single crystal surface held at \SI{300}{\celsius} was exposed to Gd vapor from an electron-beam heated tungsten crucible under ultra-high vacuum. \Acp{chGNR} were produced by the same on-surface reaction described previously \cite{Li2021} for which a (3,2)-DBQA organic precursor was evaporated from a Knudsen cell at \SI{300}{\celsius} onto the room-temperature \gdau{} surface, and then annealed to \SI{240}{\celsius} for \SI{20}{\minute} and \SI{360}{\celsius} for \SI{10}{\minute} to induce stepwise Ullmann coupling and cyclodehydrogenation, respectively. \ac{STM} and \ac{KPFM} measurements were realized at low-temperatures ($\approx\SI{5}{\kelvin}$) under ultra-high vacuum using a qPlus PtIr tip. Lock-in detection of differential conductance spectra was performed by modulation of the tunneling bias on the order of $1\,\mathrm{mV}_\mathrm{0p}$.

\section{Acknowledgments}
\acknowledgments{We would like to thank Sofia Sanz Wuhl and Thomas Frederiksen for fruitful discussions. The authors acknowledge financial support from grants PID2022-140845OBC61, PID2022-140845OBC62, and CEX2020-001038-M funded by \href{http://doi.org/10.13039/501100011033}{MICIU/AEI} and the European Regional Development Fund (ERDF, A way of making Europe), from the FET-Open project SPRING (863098), the ERC Synergy Grant MolDAM (no.~951519), and the ERC-AdG CONSPIRA (101097693) funded by the European Union, from project 2023-QUAN-000028-01 funded by the Diputación Foral de Gipuzkoa, and from the Xunta de Galicia (Centro singular de investigación de Galicia accreditation 2019-2022, ED431G 2019/03 and Oportunius Program).
L. E. and T.S. acknowledge  funding by the Spanish Ministerio de Educación y Formación Profesional through the PhD scholarship No. PRE2020-094813 and PRE2022-000247, respectively.
F.R.-L. acknowledges funding by the Spanish Ministerio de Educación y Formación Profesional through the PhD scholarship No. FPU20/03305. F.S. acknowledges funding by the Spanish Ministerio de Ciencia, Innovación y Universidades through the Ramón y Cajal Fellowship RYC2021-034304-I.
}

\bibliographystyle{apsrev4-1}
\bibliography{main.bib}

\end{document}